\documentclass[11pt, preprint, aps, floats, nofootinbib, preprintnumbers, groupedaddress, superscriptaddress,floatfix]{revtex4-1}

\usepackage{amsmath,amssymb,epsfig,color,bbold}
\usepackage{feynmp-auto}
\usepackage[hidelinks]{hyperref}
\usepackage{caption}
\usepackage{subcaption}
\usepackage{slashbox}
\usepackage{physics}
\usepackage{psfrag}
\usepackage{graphicx,xspace}
\usepackage{bm}

\usepackage[capitalise]{cleveref}
\usepackage{mathrsfs}  

\newcommand{\be}{\begin{equation}}  
\newcommand{\ee}{\end{equation}}

\long\def\symbolfootnote[#1]#2{\begingroup%
\def\thefootnote{\fnsymbol{footnote}}\footnote[#1]{#2}\endgroup}

\def\dd{\mathrm{d}} 

\def\iu{\mathrm{i}}

\allowdisplaybreaks

\begin{document}

\begin{fmffile}{feynmffile} 
\fmfcmd{%
vardef middir(expr p,ang) = dir(angle direction length(p)/2 of p + ang) enddef;
style_def arrow_left expr p = shrink(.7); cfill(arrow p shifted(4thick*middir(p,90))); endshrink enddef;
style_def arrow_left_more expr p = shrink(.7); cfill(arrow p shifted(6thick*middir(p,90))); endshrink enddef;
style_def arrow_right expr p = shrink(.7); cfill(arrow p shifted(4thick*middir(p,-90))); endshrink enddef;}

\fmfset{arrow_ang}{15}
\fmfset{arrow_len}{2.5mm}
\fmfset{decor_size}{3mm}
\begin{titlepage}

\begin{flushright}
CALT-TH/2024-007\\
\today
\end{flushright}

\vspace{0.7cm}
\begin{center}
\Large\bf 
Generalized eikonal identities for charged currents
\end{center}

\vspace{0.8cm}
\begin{center}
{\sc   Ryan Plestid }\\
\vspace{0.4cm}

{\it Walter Burke Institute for Theoretical Physics,\\
California Institute of Technology, 
Pasadena, CA, 91125 USA\vspace{1.2mm}}
\end{center}
\vspace{1.0cm}
\begin{center}
\today
\end{center}

\vspace{1.0cm}

  \vspace{0.2cm}
  \noindent
    We discuss QED radiative corrections to contact operators coupling two heavy fields and one light field. These operators appear ubiquitously in weak interactions with nuclei such as beta decay and neutrino nucleus scattering.  New eikonal identities are derived in the static limit (i.e., neglecting nuclear recoil) that allow for manifest power counting of enhancements proportional to the charge of the nucleus. We apply these new identities to nuclear beta decays and find that the ``independent particle model'' used by Jaus, Rasche, Sirlin \& Zucchini is closely related, though not identical, to a model independent effective field theory calculation.

\end{titlepage}

\tableofcontents
\vfill 
\pagebreak
\newpage

\section{Introduction \label{Intro}}

The heavy particle limit of gauge theories is dramatically simplified by eikonal identities \cite{Korchemsky:1991zp,Isgur:1989vq,Isgur:1990yhj,Georgi:1990um,Falk:1990yz,Bauer:2002nz,Becher:2009qa,Collins:1989gx,Grozin:2022umo}. These eikonal identities make use of the simplified limit of propagators for matter fields. For example, a scalar particle with momentum $p=M v$ put off-shell by a photon with momentum $q$ becomes
\[ 
    \frac{\iu}{(p+q)^2-m^2} \rightarrow \frac{\iu}{2 M v\cdot q}~,
\]
in the limit that $q \ll M$. The propagator on the right-hand side is linear (rather than quadratic) in $q$, and this allows for substantial simplifications.  In particular, this linear behaviour of propagators allows the use of partial fraction identities after which the sum over many permutations of photon insertions often reduces to a product of simple factors. Many otherwise intractable problems are reduced to bookkeeping and combinatorics.

Perhaps the most famous example is the result of Yennie Frautschi \& Suura (YFS) regarding the factorization of soft-radiation in QED \cite{Yennie:1961ad}. Splitting functions in QCD and QED make heavy use of eikonal algebra \cite{Gribov:1972ri,Altarelli:1977zs,Dokshitzer:1977sg}. The same eikonal properties underlie the simplifications inherent to Wilson lines \cite{Korchemsky:1991zp,Grozin:2022umo}.  A related identity allows one to demonstrate the emergence of classical background, e.g.\ Coulomb, fields sourced by heavy particles \cite{Brodsky:SLAC1010,Neghabian:1983vm,Weinberg:1995mt}. Suffice to say, eikonal algebra is a key tool in the study of soft limits for gauge theories. 

Surprisingly little is known about the application of eikonal identities in the context of charged currents.   Perhaps the most relevant example is nuclear beta decay \cite{Hardy:2020qwl} which involves a heavy-heavy-light vertex formed by a nucleus of charge $Z$, a nucleus of charge $Z\pm 1$, and a single electron/positron. The same vertex appears in charged-current neutrino nucleus (or nucleon) scattering. Precision theory for both processes are important for modern experimental programs in fundamental physics \cite{Branca:2021vis,Tomalak:2022xup,Seng:2018yzq,Czarnecki:2019mwq,Hardy:2020qwl}. More complicated scenarios are furnished in e.g.\ double-beta decay \cite{Dolinski:2019nrj} where the vertex would be heavy-heavy-light-light and the charge exchange with the hadronic system is $\pm 2$. 

In the context of hadronic weak interactions, it is crucial to understand how eikonal identities are modified in the presence of charged currents. Nucleons and nuclei can be treated using a heavy particle effective field theory (EFT) formalism. Reactions involving nuclei e.g., 
\begin{align}
    ~^{10}{\rm C} &\rightarrow  ~^{10}{\rm B} + e^+ + \nu_e ~,\\
    ~\nu_\ell ~^{40}{\rm Ar} &\rightarrow  ~^{39}{\rm Cl} + p^+ + \ell^-~,
\end{align}
involve heavy particles with electric charged $Z$ and $Z+z$ (with $z\neq 0$) in the initial and final state respectively. The transition is mediated by a contact interaction (the weak force) carrying non-zero electric charge. It is this phenomenological scenario that is the focus of this paper.

In this work we derive new eikonal identities for charged current processes. In \cref{Eikonals} we derive the relationships diagrammatically  using partial fraction identities and combinatorics. In \cref{Beta-decay} we apply these identities to nuclear beta decay and find that their application results in substantial simplifications in the analysis of QED corrections at high loop order. We identify new  gauge invariant sub-classes of diagrams that emerge in the heavy-particle limit. We use these gauge invariant sub-classes to show that the ``independent particle model'' introduced by Jaus and Rasche \cite{Jaus:1970tah,Jaus:1972hua} and used by Sirlin and Zuchini \cite{Sirlin:1986cc,Sirlin:1986hpu} for outer radiative corrections in beta decay, is nearly (but not exactly) equivalent to a model independent EFT calculation. We conclude in \cref{Conclusions} with a discussion of potential future applications.

\section{Eikonal identities for charged currents \label{Eikonals} }
In this section we will analyze diagrams involving $n$ (virtual or real) photons, an incoming heavy particle, $A$, of charge $Z$, and an outgoing heavy particle, $B$, of charge $Z+z$. The transition of $A\rightarrow B$ is mediated by an external current. All of the photon momenta $\{q_i\}$ are taken small compared to the heavy particle mass scale. Our goal is to demonstrate that these diagrams reduce to a sum of diagrams where the photons connect either to a background Coulomb field with charge $Z$, or to a heavy particle of charge $z$. In this second form, power counting in $Z$ becomes manifest and it is easy to track enhancements due to a large nuclear charge.

We focus on the limit of soft virtual photons and work in a low-energy EFT where photon wavelengths are long compared to the nuclear radius. This allows particles $A$ and $B$ to be treated as point-like heavy particles carrying four-velocity labels $v_A$ and $v_B$ \cite{Manohar:2000dt}.   Heavy particles have eikonal propagators, $1/(v\cdot p + \iu 0)$, and are  minimally coupled to the photon field, $\mathcal{L} \supset h_v^\dagger v\cdot D h_v$ with $D_\mu = \partial_\mu -\iu Q e A_\mu$ where $Q$ is the particle's charge. We will neglect nuclear recoil, and work in the static limit where $v_A=v_B=v$; this is an excellent approximation for realistic kinematics.

Consider an external current $\mathcal{J}$ which induces the charge-changing reaction  $A\rightarrow B$. We will be interested in correlators involving $n$ photons and one external current $\mathcal{J}$.  As a concrete example, we may take $A=\!~^{10}{\rm C}$ and $B=\!~^{10}{\rm B}$ with $\mathcal{J}$ the weak charged current. If we represent $\mathcal{J}$ by a square, the parent, $A$, with a double line, and  the daughter state, $B$ with a dashed double line, then for two photons we are interested in the elastic correlator,
\vspace{6pt}
\begin{equation}
    \label{G-example}
  G_{\mu_1\mu_2}^{(\rm el)}(q_1,q_2)= 
  \raisebox{-12pt}{
    \begin{fmfgraph*}(75,45) 
    \fmfbottom{i1,d1,o1}
    \fmftop{i2,d2,o2}
    \fmf{double}{i1,b1,b2,b3}
    \fmf{dbl_dashes}{b3,b4,b5,o1}    \fmf{phantom}{o2,t5,t4,t3,t2,t1,i2}
    \fmfv{label=$q_1$}{t2}
    \fmfv{label=$q_2$}{t4}
    \fmf{photon,tension=0}{b2,t2}
    \fmf{photon,tension=0}{b4,t4}
    \fmfv{d.sh=square,d.si=3mm}{b3}
    \end{fmfgraph*}
    }
    + ~{\rm permutations}~,
\end{equation}
More generally,  $G_{\mu_1, ..., \mu_n}^{(\rm el)}(q_1,...,q_n)$ denotes the sum over all $n$-photon dressings of the bare matrix element $\mathcal{J}_{AB}=\mel{B}{\mathcal{J}}{A}$. Different graphs which contribute to \cref{G-example} are proportional to $Z^2$, $Z(Z+z)$, or $(Z+z)^2$ [{\it cf.} \cref{ZA_Sq,ZA_ZB,ZB_Sq}]. For applications involving weak interactions with nuclei it is important to be able to systematically isolate contributions enhanced by $Z$; our goal is to make power-counting with $Z$ manifest. 

We may simplify our analysis considerably by using the standard soft-photon identity \cite{Yennie:1961ad,Weinberg:1965nx}
\begin{equation}
    \sum_{\rm perms} \frac{1}{v\cdot q_1+\iu0} \frac{1}{v\cdot (q_1+q_2)+\iu0} \times \hdots\times  \frac{1}{v\cdot (\sum_{i=1}^n q_i)+\iu0}  = \prod_{i=1}^n \frac{1}{v\cdot q_i +
    \iu 0} ~. 
\end{equation}
We may partition the set of crossed ladders into those with no photons to the right $\mathcal{J}_{AB}$, one photon to the right of $\mathcal{J}_{AB}$ etc. Let us introduce the set $\mathcal{N}=\{1,2,...,n\}$ and the set $\mathcal{N}_{i,j,k} = \mathcal{N}/\{i,j,k\}$ (i.e.\ the set ``not $i$, $j$, or $k$''). For a set of integers $S$ let us denote the symmetrized product of soft photon emissions from the initial state by $L(S)$ and by the final state by $R(S)$, i.e.\ 
\begin{align}
    L(S)&= \prod_{i\in S}^n \frac{1}{v\cdot q_i+\iu 0} ~,\\
    R(S)&= \prod_{i\in S}^n \frac{1}{-v\cdot q_i+\iu 0} ~.
\end{align}
We may then write the result for general $z$ as (with $i,j\in \mathcal{N}$)
\begin{equation} 
    \begin{split}
    G^{(\rm el)}_{\mu_1,...,\mu_n;\nu}(q_1,...,q_n) = \qty[\prod_{i} v_{\mu_i} ] \times \bigg[ &
    (Z+z)^n R(\mathcal{N}) \\
    &+ Z(Z+z)^{n-1}\sum_{i} R(\mathcal{N}_i) L(\{i\})  \\ 
    & + Z^2(Z+z)^{n-2}\sum_{i>j} R(\mathcal{N}_{i,j}) L(\{i,j\})\\
    & + \ldots  \\
    &+Z^{n-1}(Z+z) \sum_{i} R(\{ i\}) L(\mathcal{N}_i)\\
    &+Z^n L(\mathcal{N}) \bigg]~.
    \end{split}
\end{equation}
This can be written diagrammatically using a square for the external current, drawing the parent, $A$, with a double line, and drawing the daughter state, $B$ with a dashed double line. The resulting diagrams at two-loop order are given by, 
\begin{align}
\label{ZA_Sq}
O(Z_A^2) = 
\raisebox{-20pt}{
\begin{fmfgraph*}(75,45) 
    \fmfbottom{i1,d1,o1}
    \fmftop{i2,d2,o2}
    \fmf{double}{i1,b1,b2,b3}
    \fmf{dbl_dashes}{b3,b4,b5,o1}
    \fmf{phantom}{o2,t5,t4,t3,t2,t1,i2}
    \fmf{photon,tension=0}{b1,t1}
    \fmf{photon,tension=0}{b2,t2}
    \fmfv{d.sh=square,d.si=3mm}{b3}
\end{fmfgraph*}
}
\quad + \quad 
\raisebox{-20pt}{
\begin{fmfgraph*}(75,45) 
    \fmfbottom{i1,d1,o1}
    \fmftop{i2,d2,o2}
    \fmf{double}{i1,b1,b2,b3}
    \fmf{dbl_dashes}{b3,b4,b5,o1}
    \fmf{phantom}{o2,t5,t4,t3,t2,t1,i2}
    \fmf{photon,tension=0}{b1,t2}
    \fmf{photon,tension=0}{b2,t1}
    \fmfv{d.sh=square,d.si=3mm}{b3}
\end{fmfgraph*}
}
~,\\
\label{ZA_ZB}
O(Z_A Z_B) = 
\raisebox{-20pt}{
\begin{fmfgraph*}(75,45) 
    \fmfbottom{i1,d1,o1}
    \fmftop{i2,d2,o2}
    \fmf{double}{i1,b1,b2,b3}
    \fmf{dbl_dashes}{b3,b4,b5,o1}    \fmf{phantom}{o2,t5,t4,t3,t2,t1,i2}
    \fmf{photon,tension=0}{b1,t1}
    \fmf{photon,tension=0}{b5,t5}
    \fmfv{d.sh=square,d.si=3mm}{b3}
\end{fmfgraph*}
}
\quad+ \quad
\raisebox{-20pt}{
\begin{fmfgraph*}(75,45) 
    \fmfbottom{i1,d1,o1}
    \fmftop{i2,d2,o2}
    \fmf{double}{i1,b1,b2,b3}
    \fmf{dbl_dashes}{b3,b4,b5,o1}
    \fmf{phantom}{o2,t5,t4,t3,t2,t1,i2}
    \fmf{photon,tension=0}{b1,t5}
    \fmf{photon,tension=0}{b5,t1}
    \fmfv{d.sh=square,d.si=3mm}{b3}
\end{fmfgraph*}
}
~,\\
\label{ZB_Sq}
O(Z_B^2) = 
\raisebox{-20pt}{
\begin{fmfgraph*}(75,45) 
    \fmfbottom{i1,d1,o1}
    \fmftop{i2,d2,o2}
    \fmf{double}{i1,b1,b2,b3}
    \fmf{dbl_dashes}{b3,b4,b5,o1}    \fmf{phantom}{o2,t5,t4,t3,t2,t1,i2}
    \fmf{photon,tension=0}{b4,t4}
    \fmf{photon,tension=0}{b5,t5}
    \fmfv{d.sh=square,d.si=3mm}{b3}
\end{fmfgraph*}
}
\quad + \quad 
\raisebox{-20pt}{
\begin{fmfgraph*}(75,45) 
    \fmfbottom{i1,d1,o1}
    \fmftop{i2,d2,o2}
    \fmf{double}{i1,b1,b2,b3}
    \fmf{dbl_dashes}{b3,b4,b5,o1}    \fmf{phantom}{o2,t5,t4,t3,t2,t1,i2}
    \fmf{photon,tension=0}{b4,t5}
    \fmf{photon,tension=0}{b5,t4}
    \fmfv{d.sh=square,d.si=3mm}{b3}
\end{fmfgraph*}
}
~.
\end{align}
Setting $z=0$ we obtain the standard result in the static limit \cite{Brodsky:SLAC1010,Dittrich:1970vv,Neghabian:1983vm,Weinberg:1995mt}, 
\begin{equation} 
    \begin{split}
      \bigg[ R(\mathcal{N}) + \sum_{i\in \mathcal{N}} R(\mathcal{N}_i) &L(\{i\})
        + \sum_{i\in \mathcal{N}}\sum_{j\in \mathcal{N}_i} R(\mathcal{N}_{i,j}) L(\{i,j\})\\
    & + \ldots  + \sum_{a\in \mathcal{N}} R(\{ a\}) L(\mathcal{N}_a)+Z^n L(\mathcal{N})\bigg] = \prod_{i\in \mathcal{N}} (2\pi\iu) \delta(v\cdot q_i)~.
    \end{split}
\end{equation}
The general result can be organized in a series in $Z^{n-m} z^{m}$ for $0\leq m\leq n$. The $Z^n$ contributions match the equal charge limit, and are given by the expressions presented above. Let us consider the contributions proportional to $Z^{n-1}z$. We pick up a binomial coefficient from the expansion of $(Z+z)^n$ which we will write explicitly as ${n \choose 1}$. We then have
\begin{equation}
    \label{dummy-sum}
    \begin{split}
    Z^{n-1} z \bigg[ {n \choose 1} R(\mathcal{N}) + {n-1\choose 1} \sum_{i} R(\mathcal{N}_i) &L(\{i\})  + {n-2 \choose 1}\sum_{i>j} R(\mathcal{N}_{i,j}) L(\{i,j\})\\
    & \hspace{0.1\linewidth}+ \ldots  + {1\choose 1}  \sum_{i} R(\{ i\}) L(\mathcal{N}_i)\bigg]~.
    \end{split}
\end{equation}
The binomial factors can all be reproduced  by introducing an additional ``dummy sum'' to each term, i.e.\ ${n \choose 1} R(\mathcal{N}) = \sum_{a\in \mathcal{N}} R(\mathcal{N})$ and ${n-1\choose 1} \sum_{i\in \mathcal{N} } R(\mathcal{N}_i)L(\{i\}) = \sum_{i}\sum_{a \in  \mathcal{N}_i} R(\mathcal{N}_i)L(\{i\}) $. The indices of the original sum in each term of \cref{dummy-sum} are always excluded from the second sum. This gives us 
\begin{equation}
    \begin{split}
    &Z^{n-1} z  \bigg[ \sum_a R(\mathcal{N})  + \sum_{i} \sum_{a\neq i} R(\mathcal{N}_{i})L(\{i\}) +  \sum_{i>j } \sum_{a\neq i,j} R(\mathcal{N}_{i,j})L(\{i,j\}) + \ldots \bigg]~.
    \end{split}
\end{equation} 
Using the property that for $a\in S$ we have $R(S)= \frac{1}{v\cdot q_a +\iu 0} R(S/\{a\})$ we obtain 
\begin{equation}
    \begin{split}
    &Z^{n-1} z \bigg[ \sum_a \frac{1}{v\cdot q_a +\iu 0} R(\mathcal{N}_a)  + \sum_{i} \sum_{a\neq i} \frac{1}{v\cdot q_a +\iu 0} R(\mathcal{N}_{a,i})L(\{i\}) \\
    &\hspace{0.4\linewidth}+  \sum_{i>j } \sum_{a\neq i,j} \frac{1}{v\cdot q_a +\iu 0} R(\mathcal{N}_{a,i,j})L(\{i,j\}) + \ldots \bigg]~,\\
    &=Z^{n-1} z \sum_a \frac{1}{v\cdot q_a +\iu 0} \bigg[ R(\mathcal{N}_a)  + \sum_{i} R(\mathcal{N}_{a,i})L(\{i\}) \\
    &\hspace{0.4\linewidth}+  \sum_{i>j} R(\mathcal{N}_{a,i,j})L(\{i,j\}) + \ldots \bigg]~,\\
    &= Z^{n-1}z \sum_{a\in \mathcal{N}} \frac{1}{v\cdot q_a +\iu 0} \prod_{i\in \mathcal{N}_a} (2\pi \iu) \delta(v\cdot q_a)~,
    \end{split}
\end{equation} 
where in the second line the sums over $i$ and $j$ are for $i,j\in \mathcal{N}_a$. This may be recognized as the Feynman rules for $n-1$-photons coupling to a Coulomb field, and one photon coupling to a heavy-particle of charge $z$. 

The analysis presented above (for $z^1 Z^{n-1}$) generalizes readily to $z^m Z^{n-m}$ with ${n \choose 1}$ replaced by ${n \choose m}$. The dummy sum that must be introduced is appropriately modified; for example ${n \choose 2}= \sum_{i>j}1$ and ${n \choose 3} = \sum_{i>j>k} 1$. The rest of the analysis proceeds identically. The final result is that 
\begin{equation}  
    \label{main-result}
    \begin{split}
    G^{(\rm el)}_{\mu_1,...,\mu_n;\nu}(q_1,...,q_n) &= \qty[\prod_{i=1}^n v_{\mu_i}] \bigg[Z^n \prod_{i=1}^n (2\pi \iu) \delta(v\cdot q_i) \\
    &\hspace{0.15\linewidth}+ z Z^{n-1} \sum_a \frac{1}{v\cdot q_a+\iu 0} \prod_{i\neq a} (2\pi \iu)\delta(v\cdot q_i)  \\
    & \hspace{0.15\linewidth}+ z^2 Z^{n-2} \sum_{a<b} \frac{1}{v\cdot q_a+\iu 0}\frac{1}{v\cdot q_b+\iu 0} \\
    &\hspace{0.42\linewidth} \times \prod_{i\neq a,b} (2\pi \iu) \delta(v\cdot q_i) \\
    & \hspace{0.15\linewidth}+\hspace{0.1\linewidth} \ldots \hspace{0.35\linewidth}\bigg]
    \end{split}
\end{equation}
This is equivalent to the coherent sum of amplitudes from a heavy particle of charge $z$ transitioning to a heavy particle with vanishing charge and a static background Coulomb field  $V(\vb{r})=Z\alpha/|\vb{r}|$. The static Coulomb field couples to all charged particles in the diagram {\it except not} to the heavy particle of charge $z$. The sum over $a$ and $b$ accounts for crossed diagrams between Coulomb modes and the soft photons emitted by the initial state heavy particle. 

This result may have been expected \cite{Szafron_Private} on the basis on the Abelian exponentiation theorem \cite{Yennie:1961ad,Grozin:2022umo}. Since the webs for a particle of charge $Z$ and $Z+z$ will be linear in the charge such that the product of the Wilson line soft-functions would be given be proportional to $Z$. This neglects the subtlety that accounts for the Coulomb field in the $z=0$ limit, and the analysis above is a direct demonstration via combinatorics that the intuition from the exponentiation theorem is indeed correct.

\section{Effective theory of beta decay in the point-like limit \label{Beta-decay} } 
We now apply \cref{main-result} to superallowed (i.e., $0^+\rightarrow 0^+$) beta decays. These transitions currently provide the most precise extraction of the Cabibbo–Kobayashi–Maskawa matrix element $|V_{ud}|$ \cite{Hardy:2020qwl}. To extract fundamental physics from these decays one requires control over QED radiative corrections at high loop order. Since nuclei in these transitions have charge much greater than unity (ranging from $Z=6$ for $^{10}{\rm C}$ to $Z=37$ for $^{74}{\rm Rb}$) it is important to systematically include $Z$-enhanced radiative corrections. In what follows we sketch how the eikonal identities derived above can be used to dramatically reduce the number of diagrams that must be evaluated, and to isolate $O(Z^2\alpha^3)$ corrections to superallowed beta decays. 

\subsection{Point-like effective field theory}
Let us now consider beta decay in an EFT where nuclei appear as point-like heavy particles.  In the EFT we have two heavy particle fields, $h_A$ and $h_B$, and two relativistic fermions, $e$ and $\nu$. The relevant Lagrangian is given by 
\begin{equation}
    \label{L-hpet}
    \begin{split}
    \mathcal{L}_{\rm HPET} &= -\frac14 F_{\mu\nu}F^{\mu\nu} + h_B^\dagger\qty[\iu v_\mu(\partial^\mu+\iu Z_B A^\mu)] h_B + h_A^\dagger\qty[\iu v_\mu (\partial^\mu + \iu Z_A A^\mu)]h_A   \\
    &\hspace{0.075\linewidth}+ \bar{\nu} \iu \gamma_\mu \partial^\mu \nu+ \bar{e} \qty[\iu\gamma_\mu(\partial^\mu-\iu e A^\mu)-m ]e +\qty[ -G_F \bar{e}\Gamma_\ell \nu  h_B^\dagger \Gamma_h h_A + {\rm c.c.}]~,
    \end{split} 
\end{equation} 
where $\Gamma_\ell$ and $\Gamma_h$ are the relevant spin structures for the weak charged current.

A decoupling transformation \cite{Bauer:2001yt,Becher:2014oda} can be performed on \cref{L-hpet} by introducing a field redefinition in terms of Wilson lines. The key is to shift $h_A$ and $h_B$ to 
\begin{align}
    h_A\rightarrow \mathfrak{h}_A(x) &= S(x) h_A(x)  ~,\\
    h_B\rightarrow \mathfrak{h}_B(x) &= \bar{S}(x)h_B(x)~.
\end{align}
where $S(x)$ and $\bar{S}(x)$ are Wilson lines appropriate for particles in the initial and final state with charge $Ze$, 
\begin{align}
    S(x) &= \exp\qty[ \iu Z e \int_{-\infty}^0 \dd s ~v\cdot A(x+sv) ]~.\\
    \bar{S}(x)&= \exp\qty[- \iu Z e \int_0^{\infty} \dd s ~v\cdot A(x+sv) ],
\end{align}
In terms of the new fields, and using $z=Z_B-Z_A$, the Lagrangian assumes the form, 
\begin{equation}
    \label{L-hpet-decoupled}
    \begin{split}
    \mathcal{L}_{{\rm HPET}'} &= -\frac14 F_{\mu\nu}F^{\mu\nu} + \mathfrak{h}_B^\dagger\qty[\iu v_\mu(\partial^\mu)] \mathfrak{h}_B + \mathfrak{h}_A^\dagger\qty[\iu v_\mu (\partial^\mu + \iu z A^\mu)]\mathfrak{h}_A   \\
    &\hspace{0.075\linewidth}+ \bar{\nu} \iu \gamma_\mu \partial^\mu \nu+ \bar{e} \qty[\iu\gamma_\mu(\partial^\mu-\iu e A^\mu)-m ]e +\qty[ -G_F \bar{e}\Gamma_\ell \nu  \mathfrak{h}_B^\dagger \Gamma_h \mathfrak{h}_A S \bar{S}^\dagger + {\rm c.c.}]~,
    \end{split} 
\end{equation} 
Now background Coulomb diagrams arise from the matrix element of the Wilson lines $\langle S \bar{S}^\dagger \rangle$. The differing boundary conditions in position space for the Wilson lines $S$ and $\bar{S}$ reproduce the differing causal regulators in momentum space. This can be seen explicitly as 
\begin{equation}
    \bar{S}^\dagger(x) S(x)= \exp\qty[ \iu Z e \int_{-\infty}^\infty\dd s v\cdot A(x+sv)] = \exp[2\pi \iu \delta(v\cdot \partial) v\cdot A]~. 
\end{equation}
In the decoupled theory there is a single heavy particle, $\mathfrak{h}_A$, with residual charge $z$,  and a background Coulomb field with charge $Z$. 

Diagrams involving Coulomb exchanges with the heavy field do not contribute to amplitudes. We can see this in two different ways: First, consider working diagram by diagram, one can note that these diagrams can always be canceled by a mass counter term which enforces a vanishing residual mass for the heavy particle\footnote{The diagrams vanish in dimensional regularization unless a photon mass is included as an IR regulator. In this case the counter term enforces zero residual mass order-by-order in $Z\alpha$.} \cite{z2a3anom}. Second, one can show that these diagrams belong to a gauge invariant sub-class (see \cref{Gauge-inv-sub-class}), and that this subclass vanishes.

\begin{figure}[t]
\centering
\begin{subfigure}[]{0.25 \textwidth}
\parbox{40mm}{
\begin{fmfgraph*}(120,60)
    \fmfstraight
  \fmfleftn{l}{3}
  \fmfrightn{r}{3}
  \fmfbottomn{b}{9}
  \fmf{phantom}{l2,v,r2}
  \fmffreeze
  \fmf{fermion}{r3,x,y,z,v}
  \fmf{double}{l2,w,v}
  \fmffreeze
  \fmf{photon}{x,b8}
  \fmf{photon}{y,b7}
  \fmf{photon,right}{z,w}
  \fmfv{decor.shape=cross}{b8}
  \fmfv{decor.shape=cross}{b7}
      \fmfv{decor.shape=square}{v}
\end{fmfgraph*}
}
\subcaption{$(a)$}
\end{subfigure}
\begin{subfigure}[]{0.25\textwidth}
\parbox{30mm}{
\begin{fmfgraph*}(120,60)
    \fmfstraight
  \fmfleftn{l}{3}
  \fmfrightn{r}{3}
  \fmfbottomn{b}{9}
  \fmf{phantom}{l2,v,r2}
  \fmffreeze
  \fmf{fermion}{r3,x,y,z,v}
  \fmf{double}{l2,w,v}
  \fmffreeze
  \fmf{photon}{x,b8}
  \fmf{photon}{z,b6}
  \fmf{photon,right}{y,w}
  \fmfv{decor.shape=cross}{b8}
  \fmfv{decor.shape=cross}{b6}
      \fmfv{decor.shape=square}{v}
\end{fmfgraph*}
}
\subcaption{$(b)$}
\end{subfigure}
\begin{subfigure}[]{0.25\textwidth}
\parbox{30mm}{
\begin{fmfgraph*}(120,60)
\fmfstraight
  \fmfleftn{l}{3}
  \fmfrightn{r}{3}
  \fmfbottomn{b}{9}
  \fmf{phantom}{l2,v,r2}
  \fmf{phantom}{l1,b6,b7,b8,r1}
  \fmffreeze
  \fmf{fermion}{r3,x,y,z,v}
  \fmf{double}{l2,w,v}
  \fmffreeze
  \fmf{photon}{z,b6}
  \fmf{photon}{y,b7}
  \fmf{photon,right}{x,w}
  \fmfv{decor.shape=cross}{b6}
  \fmfv{decor.shape=cross}{b7}
      \fmfv{decor.shape=square}{v}
\end{fmfgraph*}
}\subcaption{$(c)$}
\end{subfigure}
\begin{subfigure}[]{0.25\textwidth}
\vspace{5mm}
\parbox{40mm}{
\begin{fmfgraph*}(120,60)
    \fmfstraight
  \fmfleftn{l}{3}
  \fmfrightn{r}{3}
  \fmfbottomn{b}{16}
  \fmf{phantom}{l2,v1,v2,r2}
  \fmffreeze
  \fmf{fermion}{r3,x,y,z,w,v1}
  \fmf{double}{l2,v1}
  \fmffreeze
  \fmf{photon}{x,b14}
  \fmf{photon}{y,b12}
  \fmf{photon,right=1.5}{z,w}
  \fmfv{decor.shape=cross}{b14}
  \fmfv{decor.shape=cross}{b12}
      \fmfv{decor.shape=square}{v1}
\end{fmfgraph*}
}
\subcaption{$(p1)$}
\end{subfigure}
\begin{subfigure}[]{0.25\textwidth}
\vspace{5mm}
\parbox{40mm}{
\begin{fmfgraph*}(120,60)
\fmfstraight
  \fmfleftn{l}{3}
  \fmfrightn{r}{3}
  \fmfbottomn{b}{16}
  \fmf{phantom}{l2,v1,v2,r2}
  \fmffreeze
  \fmf{fermion}{r3,x,y,z,w,v1}
  \fmf{double}{l2,v1}
  \fmffreeze
  \fmf{photon}{x,b14}
  \fmf{photon}{w,b8}
  \fmf{photon,right=1.5}{y,z}
  \fmfv{decor.shape=cross}{b14}
  \fmfv{decor.shape=cross}{b8}
      \fmfv{decor.shape=square}{v1}
\end{fmfgraph*}
}
\subcaption{$(p2)$}
\end{subfigure}
\begin{subfigure}[]{0.25\textwidth}
\vspace{5mm}
\parbox{40mm}{
\begin{fmfgraph*}(120,60)
\fmfstraight
  \fmfleftn{l}{3}
  \fmfrightn{r}{3}
  \fmfbottomn{b}{16}
  \fmf{phantom}{l2,v1,v2,r2}
  \fmffreeze
  \fmf{fermion}{r3,x,y,z,w,v1}
  \fmf{double}{l2,v1}
  \fmffreeze
  \fmf{photon}{x,b14}
  \fmf{photon}{z,b10}
  \fmf{photon,right}{y,w}
  \fmfv{decor.shape=cross}{b14}
  \fmfv{decor.shape=cross}{b10}
      \fmfv{decor.shape=square}{v1}
\end{fmfgraph*}
}
\subcaption{$(v1)$}
\end{subfigure}
\begin{subfigure}[]{0.25\textwidth}
\vspace{5mm}
\parbox{40mm}{
\begin{fmfgraph*}(120,60)
\fmfstraight
  \fmfleftn{l}{3}
  \fmfrightn{r}{3}
  \fmfbottomn{b}{16}
  \fmf{phantom}{l2,v1,v2,r2}
  \fmffreeze
  \fmf{fermion}{r3,x,y,z,w,v1}
  \fmf{double}{l2,v1}
  \fmffreeze
  \fmf{photon}{y,b12}
  \fmf{photon}{w,b8}
  \fmf{photon,right}{x,z}
  \fmfv{decor.shape=cross}{b12}
  \fmfv{decor.shape=cross}{b8}
      \fmfv{decor.shape=square}{v1}
\end{fmfgraph*}
}
\subcaption{$(v2)$}
\end{subfigure}
\begin{subfigure}[]{0.25\textwidth}
\vspace{5mm}
\parbox{40mm}{
\begin{fmfgraph*}(120,80)
  \fmfleftn{l}{3}
  \fmfrightn{r}{3}
  \fmfbottomn{b}{10}
  \fmf{phantom}{l2,v1,v2,r2}
  \fmffreeze
  \fmf{fermion}{r3,x,y,v1}
  \fmf{double}{l2,v1}
  \fmffreeze
  \fmf{photon}{x,b8}
  \fmf{photon}{y,a1}
  \fmf{photon}{a2,b6}
  \fmf{fermion,left,tension=0.5}{a1,a2,a1}
  \fmfv{decor.shape=cross}{b8}
  \fmfv{decor.shape=cross}{b6}
      \fmfv{decor.shape=square}{v1}
\end{fmfgraph*}
}
\subcaption{$(b1)$}
\end{subfigure}
\begin{subfigure}[]{0.25\textwidth}
\vspace{5mm}
\parbox{40mm}{
\begin{fmfgraph*}(120,80)
  \fmfleftn{l}{3}
  \fmfrightn{r}{3}
  \fmfbottomn{b}{10}
  \fmf{phantom}{l2,v1,v2,r2}
  \fmffreeze
  \fmf{fermion}{r3,x,y,v1}
  \fmf{double}{l2,v1}
  \fmffreeze
  \fmf{photon}{y,b6}
  \fmf{photon}{x,a1}
  \fmf{photon}{a2,b8}
  \fmf{fermion,left,tension=0.5}{a1,a2,a1}
  \fmfv{decor.shape=cross}{b8}
  \fmfv{decor.shape=cross}{b6}
      \fmfv{decor.shape=square}{v1}
\end{fmfgraph*}
}
\subcaption{$(b2)$}
\end{subfigure}

\vspace{24pt}

\begin{subfigure}[]{0.25\textwidth}
\vspace{5mm}
\parbox{40mm}{
\begin{fmfgraph*}(140,60)
    \fmfstraight
  \fmfleftn{l}{3}
  \fmfrightn{r}{3}
  \fmfbottomn{b}{16}
  \fmf{phantom}{l2,v1,v2,r2}
  \fmffreeze
  \fmf{fermion}{r3,x,y,z,w,v1}
  \fmf{double}{l2,v1}
  \fmffreeze
  \fmf{photon}{y,b12}
  \fmf{photon}{z,b10}
  \fmf{photon,right}{x,w}
  \fmfv{decor.shape=cross}{b12}
  \fmfv{decor.shape=cross}{b10}
\fmfv{decor.shape=square}{v1}
\end{fmfgraph*}
}
\subcaption{$(w)$}
\end{subfigure}
    \caption{ The 10 graphs at $O(Z^2 \alpha^3)$ relevant for superallowed beta decays $A\rightarrow B e^+ \nu_e$ (evaluated explicitly in Ref.~\cite{z2a3anom}). The $\times$ represents a background Coulomb field Feynman rule and gives rise to three-dimensional Euclidean integrals. The double line represents a heavy particle with charge $z=1$.   \label{fig:threeloopdiagrams}}
\end{figure}
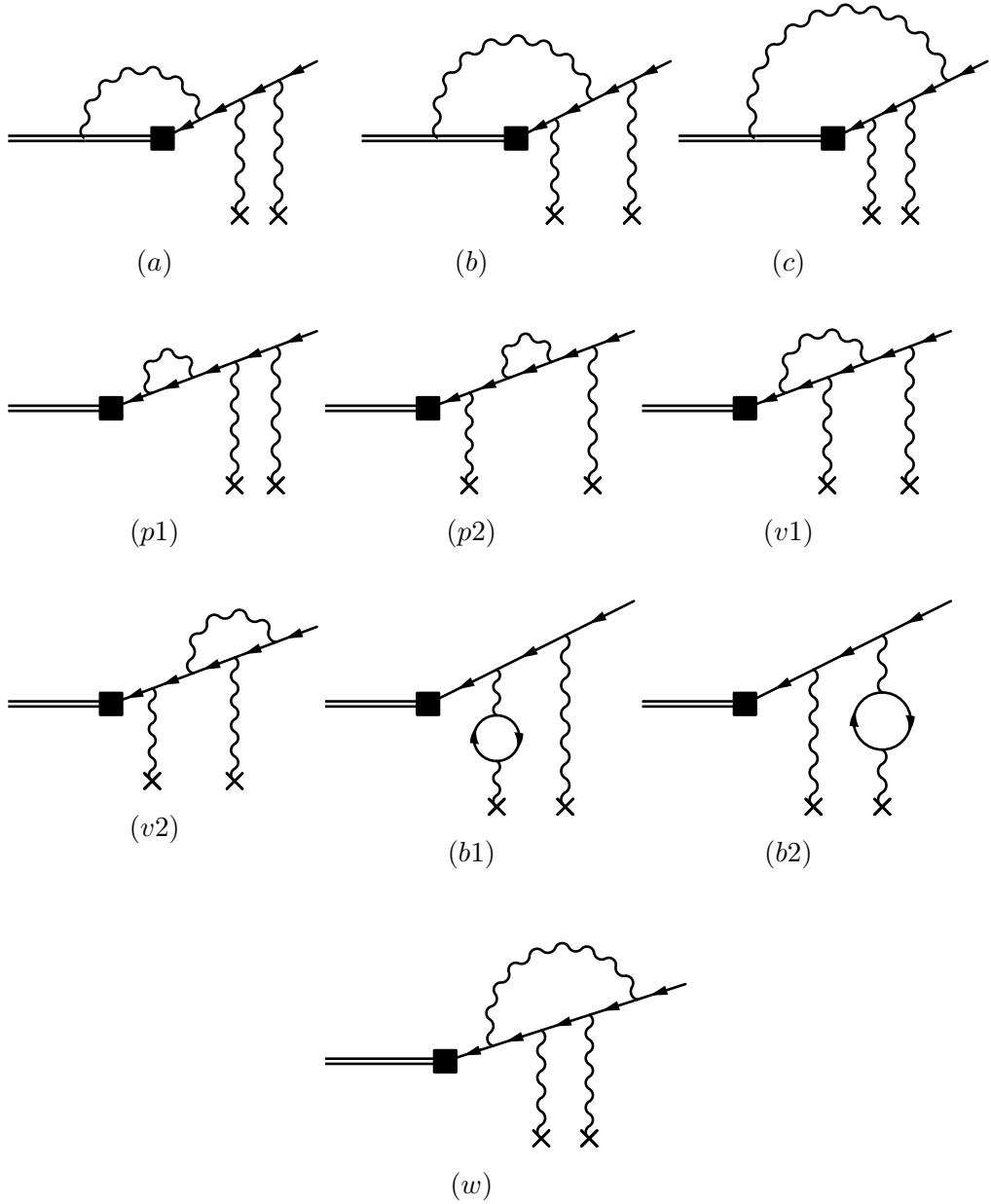

A direct evaluation of diagrams using \cref{L-hpet} is conceptually straightforward, but tedious and eventually unwieldy at high orders in perturbation theory. Current extractions of $|V_{ud}|$ from superallowed beta decay require $O(Z^2\alpha^3)$ input \cite{Sirlin:1986cc,Hardy:2020qwl}. At this order, without considering counter terms, there are 144 diagrams that would have to be evaluated. The eikonal identities presented above drastically reduces the number of diagrams which must be computed. For example at three loops one only needs to compute 10 graphs (shown in \cref{fig:threeloopdiagrams}) for the amputated amplitude using the background-field Feynman rules derived above \cite{z2a3anom}.

\subsection{Comparison with Jaus, Rasche, Sirlin \& Zucchini} 
\Cref{L-hpet} provides a model independent, and therefore universal, description of long-distance QED corrections to beta decays and other low-energy charged current processes. It is useful to compare the structure of these model independent calculations to historial approaches used for the calculation of long-distance corrections. 

As a concrete comparison let us consider long-distance (or ``outer'') radiative corrections to beta decays. Calculations are have historically been performed in the independent particle model   \cite{Jaus:1970tah,Jaus:1972hua,Sirlin:1986cc,Jaus:1986te,Sirlin:1986hpu}. This model corresponds to that defined by $\mathcal{L}_{\rm bkg}$ in \cref{L-bkg} except that in place of a heavy particle, the authors use a soft-photon or YFS approximation, making the replacement
\begin{equation}
    \frac{1}{v\cdot q + \iu 0 } \rightarrow \frac{2m_p}{2 p\cdot q + q^2 + \iu 0}~,
\end{equation}
in their diagrams. the inclusion of $q^2$ in the denominator renders the diagrams UV convergent but introduces a dependence on the ``proton mass''.  

One may treat $m_p$ as a new hard scale in the problem and separate scales using the method of regions \cite{Beneke:1997zp,Jantzen:2011nz}. The hard region supplies a contribution to the Wilson coefficients that depends on $m_p$. This dependence is unphysical, since the propagating degrees of freedom at low momenta in are the atomic nuclei of charge $Z$ and $Z\pm 1$. 
In other work by Sirlin \cite{Sirlin:1986hpu}, a charge form factor is included for the Coulomb field, which can in certain cases eliminate sensitivity to $m_p$, replacing it by the physical scale of nuclear structure. The soft region of the independent particle model reproduces  amplitudes computed with $\mathcal{L}_{{\rm HPET}'}$. We therefore conclude that, upon separating scales in the independent particle model, one will obtain amplitudes which have the correct long-distance behavior, but may contain spurious short-range contributions.


%
\section{Conclusions \label{Conclusions}} 
We have derived new eikonal identities that are relevant for problems with heavy particles whose charge is modified by an external charged current e.g., for semi-leptonic weak interactions. These heavy particles may be nuclei, nucleons, or other hadrons provided the soft-photons that appear have wavelengths that are long compared to scales of hadronic structure (e.g., the nuclear radius). The identities which give rise to Coulomb fields in the static limit are substantially modified. We have obtained a simple expression involving uncorrelated photon exchange between either a background Coulomb field, or a charge heavy particle of charge $Z_A-Z_B$. 

We have applied these identities to the EFT relevant for nuclear beta decay and identified new gauge invariant sub-classes of diagrams. The results presented above can be used to simplify calculations of the anomalous dimension and matrix elements of operators that mediate beta decays. Detailed calculations are presented elsewhere 
 \cite{short,z2a3anom}.

\section*{Acknowledgments}
\vspace{-6pt}
I thank  Richard Hill for collaboration on related projects, and I am specifically grateful for suggestions related to the decoupling transformation used in \cref{L-hpet-decoupled}. I thank Robert Szafron, Michele Papucci for useful discussions. I thank Richard Hill, Andreas Helset and Julio Para-Martinez for providing feedback on early versions of this manuscript. 

This work is supported by the Neutrino Theory Network under Award Number DEAC02-07CHI11359, the U.S. Department of Energy, Office of Science, Office of High Energy Physics, under Award Number DE-SC0011632, and by the Walter Burke Institute for Theoretical Physics.

\appendix

\section{Gauge invariant sub classes for beta decay \label{Gauge-inv-sub-class} } 

In this Appendix we use of the eikonal identities derived in \cref{Eikonals}, an auxiliary background Coulomb field Lagrangian, and certain useful properties of Coulomb gauge to identify gauge invariant sub-classes of diagrams. This simplifies the analysis of beta decay amplitudes at high perturbative order, since some of these sub-classes vanish. This section is complementary to the discussion of the decoupling transformation in \cref{L-hpet-decoupled}. 
 
\Cref{main-result} implies that the Feynman rules generated by \cref{L-hpet} for ladder graphs in which all photon attachments to the heavy composite lead to a leptonic line, can be reproduced order-by-order in perturbation theory by using the Lagrangian, 
\begin{equation}
    \label{L-bkg}
    \begin{split}
      \mathcal{L}_{\rm bkg} &= -\frac14 F_{\mu\nu}F^{\mu\nu} + h_B^\dagger\qty[\iu v_\mu\partial^\mu] h_B
      + h_A^\dagger\qty[\iu v_\mu (\partial^\mu + \iu z A^\mu)]h_A  + \bar{\nu} \iu \gamma_\mu \partial^\mu \nu \\
      &\hspace{0.075\linewidth}+ \bar{e} \qty[\iu\gamma_\mu(\partial^\mu-\iu e A^\mu- \iu Z\alpha \mathscr{A}^\mu)-m ]e
      +\qty[ -\sqrt{2}G_F \bar{e}\Gamma_\ell \nu  h_B^\dagger \Gamma_h h_A + {\rm c.c.}]~.
    \end{split} 
\end{equation} 
We will refer to these graphs as ``dynamically dressed'' in that they have a ladder skeleton but may be dressed by dynamical photons e.g.\ vertex corrections. The theory has a fixed classical background Coulomb field $\mathscr{A}^\mu(x)= \phi(x) v^\mu$ with $\phi(x)= 1/|\vb{r}|$ where $\vb{r}^\mu = x^\mu - v^\mu (v\cdot x)$. 

Amplitudes computed using \cref{L-bkg} can be written in the form
\begin{equation}
  \iu \mathcal{M}_{\rm bkg}= \sqrt{\mathcal{Z}_A(z)} \prod_{i=1}^n \sqrt{\mathcal{Z}_i} \times \qty( {\rm connected~diagrams})_{\rm bkg} 
\end{equation}
where $\mathcal{Z}_A$ is the wavefunction renormalization of the heavy field of charge $z$.  This amplitude is invariant under two separate gauge groups $U_{\rm QED}(1)\otimes U_{\rm bkg}(1)$, since the background field and gauge field are independent of one another. 

\begin{figure}
    \centering
    \includegraphics[width=0.475\linewidth]{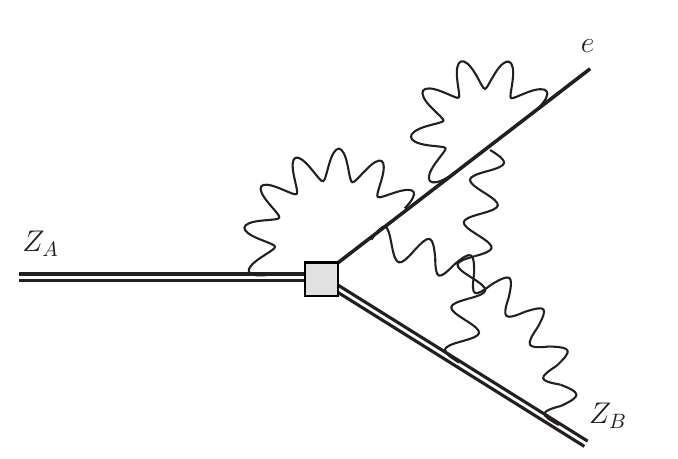}~~
    \includegraphics[width=0.475\linewidth]{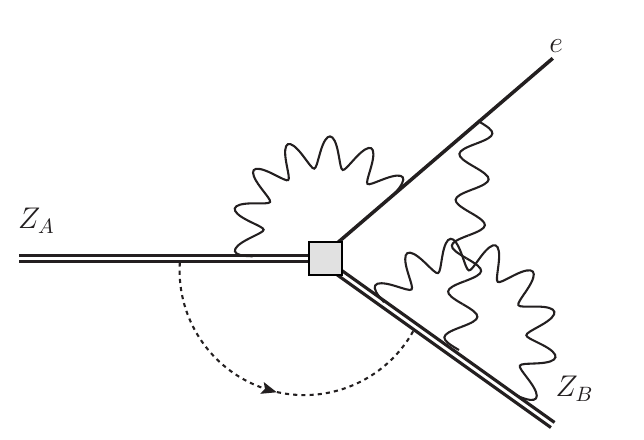}
    \caption{Examples of graphs in the dynamic subset (left) and the static subset (right). Heavy fields are drawn with double lines, while light fields (i.e.\ an electron) are drawn as single lines. The weak-current vertex is shown as a grey square (the neutrino is not drawn). The photon drawn as a dashed line on the right causes the graph to vanish in Coulomb gauge.  \label{subsets}  }
\end{figure}

Let us consider an amplitude computed using the Feynman rules of \cref{L-bkg}. We will group dynamically dressed graphs together and then add and subtract the wavefunction renormalization for the charge $z$ heavy particle order-by-order in perturbation theory.\!\footnote{One could equivalently appeal to the Abelian exponentiation theorem, since the product of Wilson lines in the initial and final state will multiply such that the overall effect is only proportional to $z$; we find the above argument cleaner.} An example of a graph from this subset is shown on the left of \cref{subsets}. We will refer to this subset of diagrams,   as the ``dynamic subset''. We will refer to the remaining diagrams, with the contribution from wavefunction renormalization subtracted, as the ``static subset''; an example of a graph in the static subset is shown on the right of \cref{subsets}. We will now argue that these two classes of diagrams are separately gauge invariant.

First notice that the dynamic subset computed using \cref{L-hpet} is equivalent to $\mathcal{M}_{\rm bkg}$ order-by-order in perturbation theory and therefore gauge invariant. To see this we make use of \cref{main-result}, which applies to any insertion of $n$ photons on a leptonic line which couple to hadrons (including the full sum over inequivalent permutations on the hadronic lines). \Cref{main-result} shows that these graphs reduce to a background Coulomb field, and Feynman rules for a charge $z$ heavy particle. The background field can be identified with $\mathscr{A}_\mu$ in \cref{L-bkg}, while terms proportional to $z^n$ in \cref{main-result} are generated by dynamical photons  ($A_\mu$) in \cref{L-bkg} coupling to the field $h_A$. Photons which begin and end on the lepton line are   also generated using \cref{L-bkg}. Since the sum of both subsets is gauge invariant, and the dynamic subset is gauge invariant, it follows that the static subset is separately gauge invariant. 

We now evaluate the static subset in Coulomb gauge (see \cref{Beg-thm} for a discussion). All diagrams contain either heavy-particle wavefunction renormalization, or heavy-heavy vertex corrections and these vanish diagram-by-diagram in Coulomb gauge. Since the static subset is gauge invariant, this statement is true for the sum of all diagrams for arbitrary gauge. Therefore, an amplitude computed with \cref{L-hpet} agrees order-by-order in perturbation theory with an amplitude computed using \cref{L-bkg}.

\section{Coulomb gauge \& the static limit \label{Beg-thm} }

In this Appendix we discuss how certain gauge invariant sub-classes of diagrams may be shown to vanish (diagram by diagram) in Coulomb gauge. This analysis applies in the static limit where $v_A=v_B=v$ i.e., neglecting the effects of nuclear recoil. 

It is convenient to work in the rest frame of the parent/daughter nucleus taking $v_\mu=(1,0,0,0)$. In Coulomb gauge transverse photons explicitly decouple from the heavy particles and Coulomb propagators, $D_{00} = \iu/\vb{q}^2$,  contain no energetic poles. For a heavy particle in the initial and final state any sub-graph involving a photon which connects two heavy particle lines will result in an integrand with all of its poles on one side of the complex plane. The contour can then be closed in the opposite direction, and the integral will vanish. For example, let us take the following loop graph
\vspace{3pt}
\begin{equation}
    \vspace{3pt}
  \raisebox{-25pt}{
    \begin{fmfgraph*}(75,45) 
    \fmfbottom{i1,o1}
    \fmftop{t1}
    \fmf{double}{i1,vl,t1}
    \fmf{dbl_dashes}{t1,vr,o1}
    \fmf{photon,tension=0}{vl,vr}
    \fmfv{d.sh=square,d.si=3mm}{t1}
\end{fmfgraph*}
 
}\quad  = \quad \int \frac{\dd^d L}{(2\pi)^d} \qty(\frac{1}{L_0 + \iu 0})^2  \frac{\iu}{\vb{L}^2}  \quad  = \quad  0 ~,
\vspace{3pt}
    \label{cusp}
\end{equation}
where we have performed the integral by contour deformation closing the integral in the upper-half plane. A corollary of \cref{cusp}, and that heavy particle self-energies vanish in Coulomb gauge, is that the cusp anomalous dimension vanishes at zero recoil  \cite{Grozin:2022umo}. 

The diagrams discussed above continue to vanish in the presence of dynamical fermions. Fermion loops can be performed first and do not mix longitudinal and transverse modes. As a result any graph containing a photon connecting two heavy lines will vanish. The only remaining graphs which are non-vanishing are ladder graphs, and ladder graphs dressed by sub-graphs on the electron lines. As discussed above these can be combined into the dynamic subset, and after adding wavefunction renormalization for heavy particle with charge $z$, form a gauge invariant subset. They may then be evaluated separately in whatever gauge is most convenient.

\end{fmffile} 

\vfill
\pagebreak

\bibliography{eikonal}

\end{document}